\documentclass[10pt,twocolumn,twoside]{article}

\usepackage[T1]{fontenc}
\usepackage{ae,aecompl}

\usepackage{graphicx}	
\usepackage{amsmath}	
\usepackage{amssymb}	
\usepackage{latexsym}
\usepackage{amsfonts}
\usepackage{marvosym}
\usepackage{wasysym}
\usepackage{color}
\usepackage{soul}
\usepackage{hyperref}
\usepackage{authblk}

\def\msun{{\rm\,M_\odot}}
\newcommand{\Diamondblack}{\mathbin{\rotatebox[origin=c]{-45}{$\blacksquare$}}}

\title{Quenching Global Star Formation: \\Dominance of Gravitational Shock Heating at $z<2$}

\author[1,2]{Jia Liu} 
\author[1]{Renyue Cen}
\affil[1] {Department of Astrophysical Sciences, Princeton University, Princeton, NJ 08544, USA}
\affil[2] {NSF Astronomy and Astrophysics Postdoctoral Fellow}

\begin{document}
\twocolumn[
\begin{@twocolumnfalse}
\maketitle

\begin{abstract}
We systematically study,
in the context of the standard cold dark matter model, star-formation suppression effects of two important known physical processes---photoheating due to reionization of the intergalactic medium and gravitational shock heating due to formation of massive halos and large-scale structure---on the global evolution of star formation rate (SFR) density 
and the so-called cosmic downsizing phenomenon in the redshift range $z=$~0--6.
We show that the steep decline of 
cosmic SFR density from $z\approx 2$ to $z=0$ 
can be primarily explained by gravitational shock heating in two forms: massive halo self-quenching and hot environment. 
Simultaneously, we show a decreasing trend in the average SFR of star-forming galaxies 
from $z=2$ to $z=0$, reproducing the observed cosmic downsizing at $z \le 2$.
Nevertheless, the average halo mass of star-forming galaxies is found to continue upsizing from $z=2$ to $z=0$. 
In stark contrast to $z<2$,
both photoheating and gravitational shock heating effects are found to play a minor role in suppressing star formation. 
Additional negative feedback effects
are required to reconcile our model with observations at $z>2$.
Internal feedback from stellar evolution and supermassive black hole growth are the natural candidates for this role, 
as galaxies at $z>2$ are more moderate in mass but stronger in star formation and are thus more vulnerable.
Our physical model can be used to treat star formation in cosmological N-body simulations.

\end{abstract}

\end{@twocolumnfalse}
]

\section{Introduction}
\label{sec:intro}
Observational evidence shows that star formation was most vibrant in massive galaxies at early cosmic times, and shifts to be in smaller galaxies towards the present day~\cite{2000Brinchmann,2005Juneau,2011Karim}. In other words, massive galaxies acquire the bulk of their stellar mass earlier than their less massive counterparts---the ``cosmic downsizing'' of star formation, first depicted by Ref.~\cite{1996Cowie}. This anti-hierarchical trend in star formation seems to be at odds with the ``bottom-up'' structure formation picture in the standard Lambda cold dark matter (LCDM) model. Innovative ideas have been put forth to break 
the hierarchy of galaxy formation, such as invoking internal feedback due to active galactic nuclei (AGN) to preferentially suppress star formation in more massive galaxies at lower redshift \cite{2000Kauffmann,2005Scannapieco,Croton2006,2006Bower,2008Somerville}. Another important observational fact is that the star formation rate per unit comoving volume (SFR density) has a gradual rise or displays a relatively flat trend from redshift $z\approx6$ to its peak at $z\approx$~2--3, followed by a sharp drop of about 1~dex till $z=0$~\cite{1996Madau,1998Madau}. 
 
Here we revisit these two issues {\it jointly} for the first time,
invoking well known external physical processes that can currently be reasonably quantified with confidence. We consider external baryonic physical processes that impede efficient cooling and/or cold gas accretion onto galaxies. Three processes are considered.
First, cosmological reionization photoheats gas to a temperature of about $10^4$K, raising the entropy of cosmic gas. As a result, halos of virial velocities below 20-50 km/s can no longer efficiently accrete gas. This physical process has suppression effect on overall SFR at high redshift ($z\ge 3$). Second, below $z \approx 3$, gas heating by gravitational shocks hinders star formation in galaxies above a certain halo mass, often referred to as ``halo mass quenching''~\cite{2005Keres,2006Dekel,2009Johansson}. 
Third, gravitational shock heating due to collapse of large-scale structure raises the temperature of cosmic gas~\cite{1999Cen,2001Dave}. Consequently, towards lower redshift, a progressively larger portion of the universe becomes filled with hot gas and star formation in galaxies residing in these hot environments is suppressed~\cite{2011bCen}.

We quantitatively demonstrate the effects due to these three processes separately 
and jointly on the average SFR and the global SFR density.
We show that without invoking any other process,
the gravitational shock heating effects can mostly reproduce
the observed cosmic downsizing with respect to SFR
and the decline in SFR density from $z=2$ to $z=0$.
An interesting outcome from our analysis is that,
while the mean SFR of star forming 
galaxies decreases with decreasing redshift,
the mean halo mass of these galaxies 
is still expected to increase with decreasing redshift.

There appears to exist an apparent,
significant tension between our model with external heating only and
observations at $z\ge 2$, where the former shows 
a continuous rise of SFR density up to $z\approx 4.5$ compared to 
currently observed SFR density peaking at $z\approx 2$.
This tension may be alleviated if the current observations
have significantly underestimated SFR density beyond $z\approx 2$.
Alternatively, the culprit may be on the theoretical side,
perhaps indicative of additional negative feedback from 
stellar evolution or AGN, which is not included in our treatment.
We argue that this needed ``internal'' feedback 
can be more naturally accommodated since both star formation and 
AGN activities are indeed most vigorous in the redshift
range of $z\approx$~2--4.5, in contrast to the lower redshift range
when both activities are much diminished.

The outline of the paper is as follows.
We first describe our model in section~\ref{sec:model}. 
We break down the impact on star formation from each effect, and compare the model that includes all three effects to multiple wavelength data in section~\ref{sec:results}. We discuss the implications in section~\ref{sec:discussion} and
 conclude in section~\ref{sec:conclusions}.

\section{Simulations and Physical Model}
\label{sec:model}

\subsection{Simulations}
\label{sec:sims}

The analysis performed utilizes the high-resolution Bolshoi N-body simulation\footnote{\url{http://hipacc.ucsc.edu/Bolshoi/MergerTrees.html}}~\cite{2016Klypin}. The simulation has a box size of 250 (Mpc/h)$^3$, 2048$^3$ particles, particle mass resolution 1$.5\times10^{8}$~$M_\odot$, with cosmological parameters $\Omega_m$=0.307, $\Omega_b=0.048$, $\Omega_\Lambda=0.693$, $\sigma_8=0.823$, $n_s=0.96$, $h=0.678$~\cite{2014PlanckParams}. We use halo catalogues, created and provided
by Ref.~\cite{2013Behroozi} using the ROCKSTAR code \cite{2013bBehroozi}, from 
thirty redshift snapshots between $z=$~0--6.
We implement physical processes on the halos that we describe now.

\subsection{Physics}
\label{subsec:phys}

We consider two external baryonic physical processes that impede gas accretion onto galaxies or prevent gas from cooling.
Our treatment explicitly does not invoke any internal feedback processes, 
such as from supernovae or AGN.

The first process---the photoionization and photoheating due to cosmological reionization and subsequent maintenance of it---raises the temperature of the intergalactic medium to
about $10^4$K, significantly impeding gas accretion to 
halos of virial velocities below $\approx 20$--$50$ km/s 
\cite{1996Thoul,1996Quinn, 1997Weinberg, 1997bNavarro, 2000Gnedin, 2001Barkana, 2008Okamoto}.
This physical process, as will be shown, 
has a suppression effect on the overall SFR primarily at high redshift ($z\ge 3$) when the typical halo mass
is comparable or does not significantly exceed 
the Jeans mass imposed by photoheated gas.

The second process---gravitational shock heating due to 
structure formation on large scales---may be categorized into two conceptually separate effects: the self-heating effect and the environmental effect.
For the former, the gas heating rate due to gas mass (along with dark matter) 
accretion 
exceeds the gas cooling rate in halos
more massive than a certain threshold, a process for which we give
a new, physically more self-consistent formulation for the division
between the so-called cold and hot accretion modes
\cite{2005Keres,2006Dekel} in the next subsection.
The star formation in galaxies more massive than the redshift-dependent
division mass is self-quenched, often referred to as ``halo mass quenching''.
For the latter, below $z\approx3$, 
gravitational shock heating due to collapse of large-scale structure 
raises the temperature of cosmic gas above that due to photoheating
\cite{1999Cen,2001Dave}.
As a result, in an increasingly larger fraction of mass in the universe,
primarily in the vicinity of groups and clusters of galaxies and large filaments,
gas is heated to high temperatures with long cooling times,
depriving galaxies in these regions of cold gas due to combined detrimental effects
of ram-pressure
stripping and starvation
\cite{2011bCen}.
Thus, star formation in galaxies residing in these hot environments is 
greatly suppressed or quenched, if not already halo mass self-quenched.

\subsection{A New Formulation of Halo Mass Quenching}
\label{sec:massquenching}

Above a certain halo mass, shock heating may overwhelm radiative cooling
to render a hot atmosphere \cite{2005Keres,2006Dekel}.
The exact formulation of this process often suffers from the 
ambiguity in defining the exact heating or cooling time scales.
Here, we take a different conceptual approach, with a focus on the overall energy  balance,
to rederive the self-quenching mass scale as a function of redshift.

We adopt the mass accretion rates of halos 
as a function of halo mass and redshift using the fitting formula
based on direct N-body simulations from~Ref.~\cite{2010Fakhouri}:
\begin{align}\label{eq:mdoth}
\dot{M}_h = & \,46.1 \,M_\odot \,{\rm yr}^{-1} \left( \frac{M_h}{10^{12}M_\odot} \right)^{1.1} \nonumber \\
 & \times\left(1+1.11z\right)\sqrt{\Omega_m(1+z)^3+\Omega_\Lambda}~,
\end{align}
which has a nearly linear dependence on halo mass. For a constant halo mass, $\dot{M}_h$ decreases towards low redshifts as a result of the Hubble expansion. 
In subsequent calculations, we assume the gas accretion rate is a constant fraction of the mass accretion rate for halos that can accrete gas,
$\dot{M}_g=(\Omega_b/\Omega_m)\dot{M}_h$. 
The gas heating rate due to gas  accretion, i.e., energy release
due to gas falling into the gravitational potential well through the 
virial sphere, is
\begin{align}\label{eq:edotheat}
\dot E_{\rm heat} = {3\over 2} \sigma_v^2 \left(\frac{\Omega_b}{\Omega_m}\right)\dot M_h~,
\end{align}
where 
$\sigma_v$ is the one-dimensional~(1D) velocity dispersion of the halo.
The gas cooling rate in the halo is 
\begin{align}\label{eq:edotcool}
\dot E_{\rm cool} = \int_0^{r_v} n_H(r)^2\Lambda(T_v)4\pi r^2 dr~,
\end{align}
where $n_H$ is the radius dependent hydrogen density \cite{1998Suto} and $\Lambda$ is the metallicity and temperature dependent cooling function \cite{sutherland93}. Here we require the gas metallicity as a function of halo mass.
To obtain that, we combine
the halo mass to stellar mass relation derived based on empirical evidence by Ref.~\cite{2013Rodrguez-Puebla}\footnote{There are many investigations on the halo to stellar mass relation, and we found practically no difference when we change the relation used here (which distinguishes central galaxies and satellites) to the ones found by~Ref.~\cite{2012Leauthaud}~(for central galaxies only) or by~Ref.~\cite{2013Behroozi}~(an average of all galaxies). This is because central halos dominate the population in this mass range.}
and the observed metallicity-stellar mass relation \cite{2004Tremonti}.  We assume the metallicity in the circumgalactic medium is 1/10 of that in the interstellar medium~\cite{Muratov2016}.
In Eq (\ref{eq:edotcool}) we assume a
constant temperature inside the virial radius for simplicity.

Setting $\dot E_{\rm heat} = \dot E_{\rm cool}$  
gives rise to the critical division halo mass $M_c$.
For halos more massive than $M_c$, the gas is kept at the virial temperature 
to form a volume filling hot atmosphere, i.e., in the hot accretion regime.
For halos of masses lower than $M_c$, a progressively larger
fraction of gas can not be kept at the virial
temperature, resulting in either an increasing fraction of direct cold accretion
or more cooling and condensing of the halo gas.

\begin{figure}
	\includegraphics[width=\columnwidth]{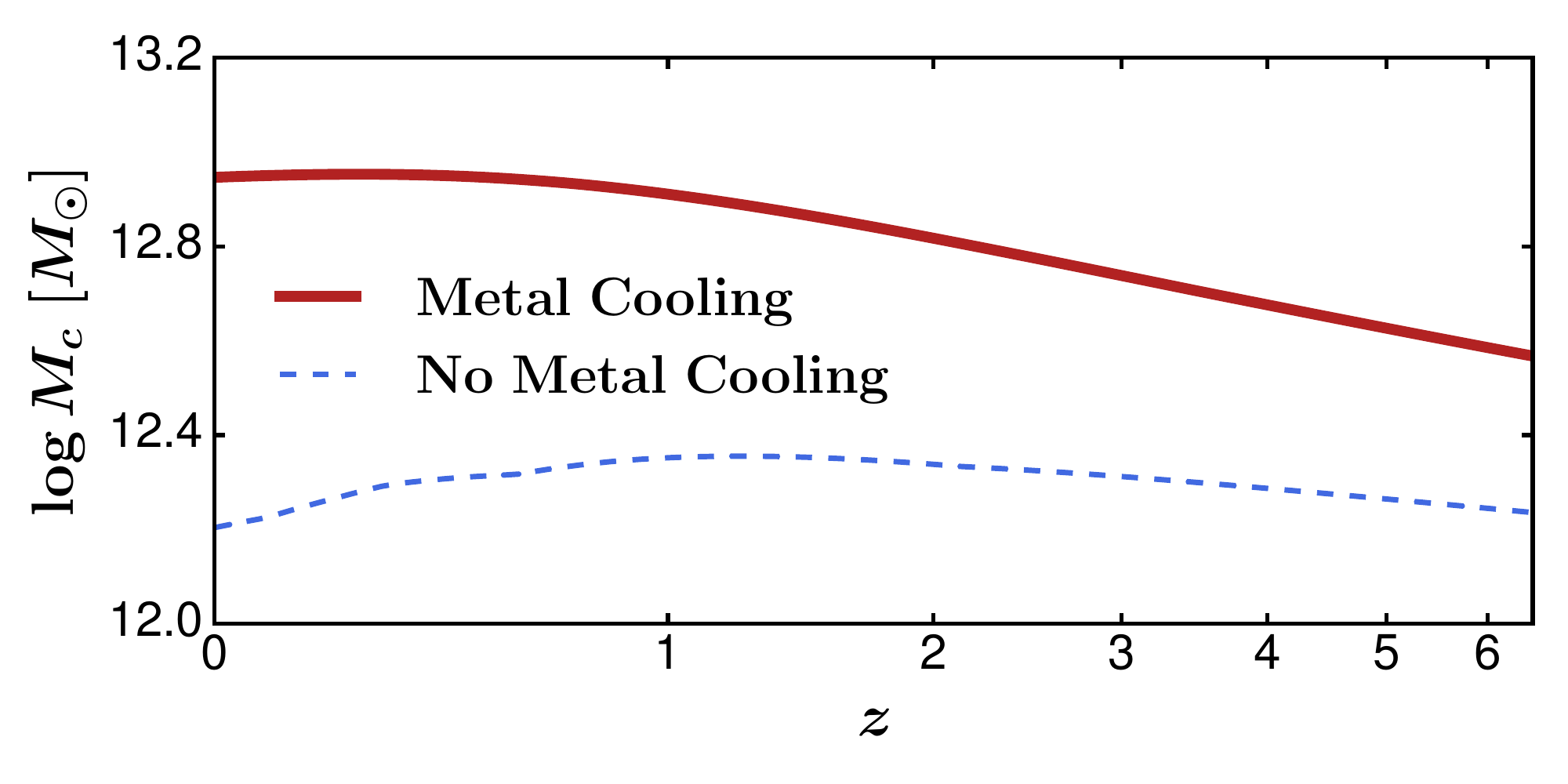}
    \vspace{-0.2in}
    \caption{The self-quenching critical mass $M_c$ as a function of redshift, calculated by setting $\dot{E}_{\rm heat}=\dot{E}_{\rm cool}$ (cf. Eqs.~\ref{eq:edotheat} and~\ref{eq:edotcool}). The gas in halos more massive than $M_c$  is kept at the virial temperature and can not form stars. }
    \label{fig:self}
\end{figure}

Fig.~\ref{fig:self} shows $M_c$ as a function of redshift.
It is noted that $M_c$
derived here with metal cooling (thick solid red curve) is significantly 
higher than that without metal cooling (thin dotted blue curve).
Ref.~\cite{2005Keres}, using detailed three-dimensional
cosmological hydrodynamic simulations without metal cooling,
derive a division mass between cold and hot accretion mode of
$M_d\approx 10^{11.4}\msun$, defined to be the halo mass 
at which the hot and cold accretion rates are equal.
Our definition of $M_c$ is where the hot accretion is 100\%.
Upon a closer examination of their results,
we find that, had our definition of $M_c$ been used, their $M_c$ would be
about $10^{12.1}\msun$ at $z=0$, in excellent agreement with our value
of $10^{12.2}\msun$ without metal cooling (thin dotted blue curve);
the small difference can be easily attributable to our assumed density profile 
or some other small details.
Interestingly, their redshift trend of $M_c$ is also consistent with our results,
with $M_c$ peaking at $z\approx 1$.
This agreement is quite reassuring, supporting both
the gas density profile and the gas accretion rate that we use in 
our derivation.

In the derivation advocated by Ref.~\cite{2006Dekel}, 
the physical argument is based on balance of cooling and 
compression of infall gas near the virial radius
where a hot halo is retained if the rate of the latter exceeds 
that of the former.
They use different but non-zero metallicity than ours
and do not use a global balance criterion that we use.
Thus, detailed comparisons can not be precisely performed,
though we note that they quote $10^{12-13}\msun$ as a possible range,
which is reasonably close to but lower than what we obtain with realistic metallicity.
It should be noted that the global cooling rate within a halo
is dominated by the central region~\cite{1995Thoul} and the effective metallicity
in our treatment is somewhat higher than that used in Ref.~\cite{2006Dekel}.
Physically, though, the compression work done by infall gas
termed in  Ref.~\cite{2006Dekel} is ultimately sourced by 
gravitational energy of the infall gas.
Thus, the statement of a global cooling and heating balance we use
is equivalent to the statement of a cooling and compression balance used
by  Ref.~\cite{2006Dekel}, except that ours is global over the entire halo,
whereas theirs is local at near the virial radius.
Since we are interested in the amount of gas cooling out to fuel star formation,
the global treatment we use is more appropriate for our purpose of 
characterizing cold gas fuel.

It is evident from Fig.~\ref{fig:self} that metal cooling 
has a major effect, increasing $M_c$ by about 0.8~dex at $z=0$
and about 0.5~dex at $z=1$ compared to 
that without metal cooling. 
We also note that halos as massive as $10^{12.5}M_\odot$ at $z=$~2--4
are typically {\it not} self-quenching.
This gives a simple, natural explanation for 
the observed high SFR of submillimeter galaxies, presumably residing in
massive galaxies at high redshift,
providing the physical basis for 
the existence of cold streams in massive halos found
in cosmological simulations \cite{2009Dekel}.

\subsection{Implementation of Physics}
\label{subsec:implementation}

We implement the three physical effects described above as follows.

First, at the low mass end, photoheating of the intergalactic gas suppresses star formation in small halos with velocity dispersion  $\sigma_v=(GM_{\rm vir}/2 r_{\rm vir})^{1/2}$ smaller than the cut-off $\sigma_c$, which is found to be in the range of 20--50~km/s~\cite{1996Thoul,1996Quinn, 2000Gnedin, 2008Okamoto}\footnote{When the circular velocity is quoted in the literature, we convert it to velocity dispersion by $\sigma_c=v_{\rm circular}/\sqrt{2}$.}.  We set SFR in halos with 1D velocity dispersion $\sigma_v < \sigma_c$ to zero.

Second, 
we set SFR in halos with masses greater than 
$M_c$ to zero, to account for 
the halo mass self-quenching, as described in section~\ref{sec:massquenching}. 
We allow a gradual transition from the fully hot accretion regime ($\log M_h>\log  M_c$) to the fully cold accretion regime ($\log M_h<\log M_c-1$), where we adopt the 1 dex transitional width from the simulation results in Ref.~\cite{2005Keres}. In particular, we allow only a fraction of accreted gas to cool, 
\begin{align}
\epsilon=0.5+0.5\tanh( 5\Delta\log M_h) \;,
\end{align} 
where $\Delta\log M_h = (\log M_c - 0.5) - \log M_h$, in halos of masses within 1 dex lower than $M_c$.
We note that this is likely due to variations in the gas density profile 
or distribution among halos at a given mass, and not due to AGN feedback, which is not implemented in their simulations.

Third, we consider the impact of hot environments---a halo that is not massive enough to self-quench may still lose its ability to form stars
due to external processes such as ram pressure stripping and starvation.
Detailed simulations present a complex physical picture of these two processes, with dependencies on a multitude of physical variables.
We adopt a simplified but relatively robust encapsulation of the physical processes. 
We assume that 
halos within a distance $d_{\rm impact}=n r_{\rm vir}$ 
of a self-quenching (i.e., $M_h>M_c$) halo of virial radius $r_{vir}$ 
 are hot gas dominated.
 We vary the parameter $n$ to examine this effect,
 although $n$ is found to be $\approx 3$,
based on insights learned from detailed simulations
\cite{2011bCen, 2014Cen} and 
consistent with observations 
~\cite{2003Gomez}.

To summarize, 
at each redshift we set the SFR to zero for halos that meet one or more of the following three conditions:

 \noindent 
(1) velocity dispersion $\sigma_v < \sigma_c$ ---photoheating effect, 

\noindent 
(2) halo mass $M_h > M_c$---self-heating (mass quenching) effect, and 

\noindent 
(3) distance from the nearby self-quenching halo $d < d_{\rm impact}$---hot environment effect.

At each redshift, 
after the removal of halos that meet the above conditions,
for halos that still remain in the star-forming category,
we compute the SFR as 
\begin{align}\label{eq:SFR}
{\rm SFR} \equiv \dot M_\star = f_{\rm int}\frac{\Omega_b}{\Omega_m}\epsilon\dot{M}_h,
\end{align}
where  $f_{\rm int}$ is the internal star formation efficiency, which
 include uncertainties  related to 
 internal baryonic processes such as gas outflows, gas accumulation, and gas recycled from stellar evolution. Throughout the rest of the paper, we set $f_{\rm int}=1$ to clearly demonstrate the suppression effects due to external feedback only. We also calculate the $f_{\rm int}(z)$ curve needed to fit the data, where we show a significantly lower value of $f_{\rm int}$ at $z>2$, possibly with a decreasing trend with increasing redshift, is required.

\section{Results}
\label{sec:results}

\subsection{Impact of External Feedback Processes}
\begin{figure*}
	\includegraphics[width=\textwidth]{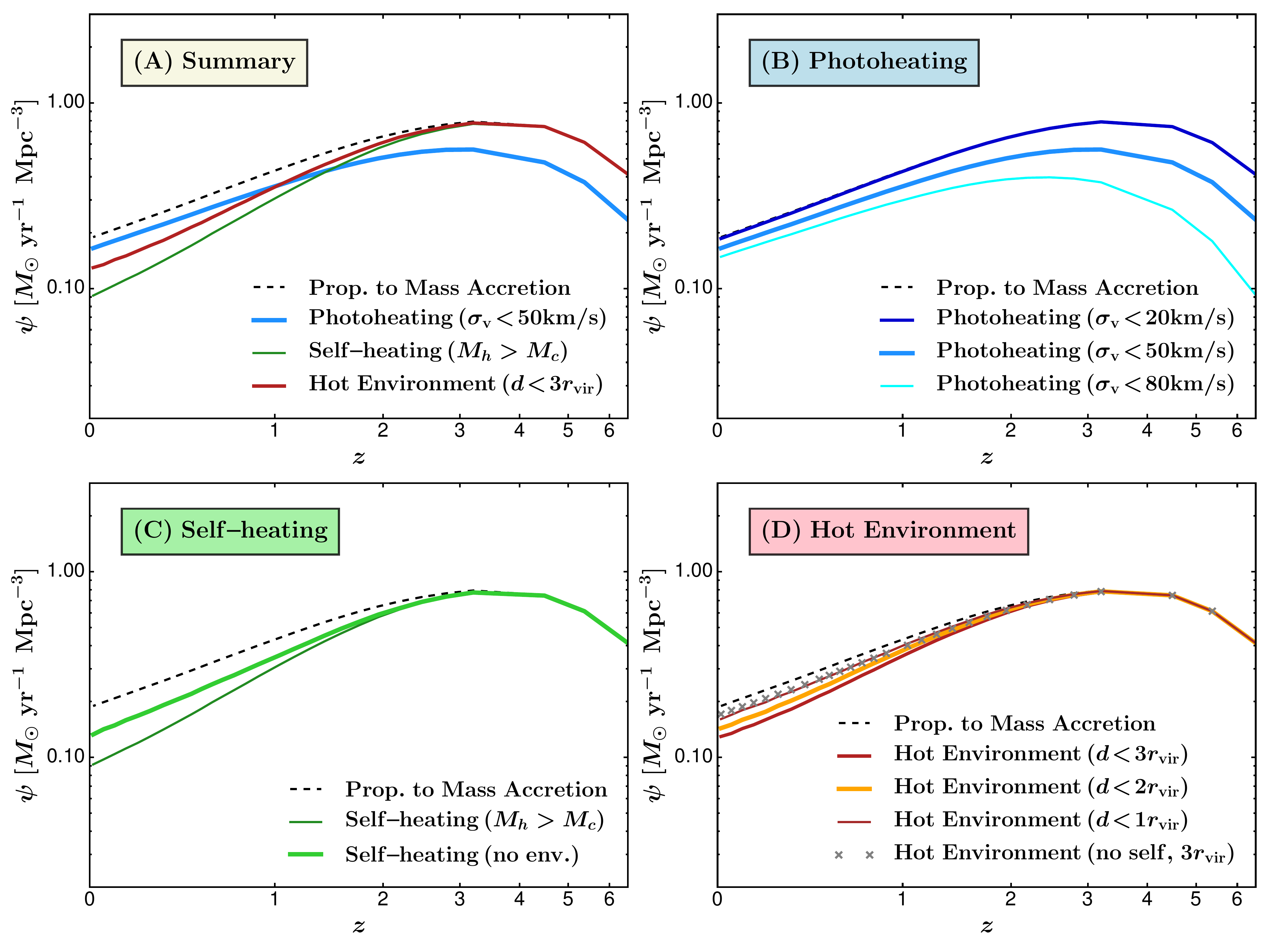}
	 \vspace{-0.2in}
    \caption{
    {\bf Panel A}: comparison of impacts from gas accretion, photoheating, self-heating, and hot environment on the star formation rate density $\psi$. It is apparent that the gas accretion rate (dashed curve) already roughly underpins the basic shape of $\psi$, with a steady increase from $z=6$ to its peak at $z=$~3--4, followed by a power law decrease towards $z=0$.  
    {\bf Panel B}: the effect of photoheating, with cut-off velocity dispersion $\sigma_c=[20,50,80]$ km/s; photoheating is more important at high redshifts where halos are small. 
    {\bf Panel C}: the impact of self-heating, which typically inhibits star formation in $\ge10^{12.5}M_\odot$ halos; because many self-heating halos are also embedded in hot environments (1/3 of them at $z=0$), we also show a comparison of cutting self-heating {\it only} halos that receive no environmental impact (thick green curve, labeled ``no env.''). 
    {\bf Panel D}: the impact of hot environments on halos around massive self-heating neighbours, through ram pressure stripping and starvation of the cold gas; $d_{\rm impact}=r_{\rm vir}$ is equivalent to satellite quenching, while the typical impact radius is estimated to be $3r_{\rm vir}$. We also show the case where we exclude halos that are also self-heating (grey crosses, labeled ``no self''). Note that in all panels we set the internal star formation efficiency $f_{\rm int}=\dot{M}_\star / \dot{M}_g=1$ to elucidate the above four effects.}
    \label{fig:sims}
\end{figure*}

In Fig.~\ref{fig:sims}, we show the impact on SFR density $\psi$ from gas accretion alone (i.e., without any external negative feedback), and with photoheating, self-heating, or hot environment, respectively and separately.
Panel (A) summarizes all four effects. First, by assuming SFR proportional to mass accretion ($\dot{M}_\star=\dot{M}_g=(\Omega_b/\Omega_m)\dot{M}_h$, black dashed line), we already see a decline in $\psi$ towards $z=0$ from the peak at $z=$~3--4.
This says that, while the overall nonlinear mass increases with decreasing redshift,
the overall rate of mass accretion onto halos 
has been steadily decreasing.
Thus, qualitatively, an overall trend of a decreasing SFR density with
decreasing redshift would be expected, even in the absence of any other physical effects.
In other words, the declining SFR density at low redshift is already underpinned,
in part, by the structure growth and Hubble expansion.
Quantitatively, however, the redshift location of the SFR density peak is too high, and the magnitude of its decline towards redshift zero is too modest from mass accretion alone.

Photoheating effect shown in Panel (B) of Fig.~\ref{fig:sims} is seen to 
suppresses the SFR in low mass halos, and has a larger impact at high redshifts where a larger fraction of collapsed mass is in
small halos and the majority of massive halos are yet to form. 
We show three levels of photoheating, with $\sigma_v$ cut-offs equal to 20, 50, and 80 km/s. It is worth noting that even if we maximize the photoheating to suppress star formation in halos with $\sigma_v<80$ km/s,  much higher than the $\leq$20 km/s limit found in the recent simulation by Ref.~\cite{2008Okamoto}, 
our model is still higher than observations by 0.2--0.3 dex at $z>3$.

The self-heating (mass quenching) effect, shown in 
Panel~(C) of Fig.~\ref{fig:sims}, 
is negligible at $z\ge 3$
but becomes increasingly important towards lower redshift, when 
the nonlinear mass increases 
and a large fraction of collapsed mass
is contained in halos more massive than $M_c$. 
The self-heating effect is seen to be strongly correlated with the hot environment effect.
The thin green solid curve in Panel~(C) is a result
of removing all halos with $M_h>M_c$,
whereas the thick green solid curve (labeled 
``no env.'') is obtained
when we only remove halos with $M_h>M_c$ that
are not in hot environments (condition \#3 in section~\ref{subsec:implementation}).
Quantitatively, the reduction of SFR density from the
black dashed curve and the (thin, thick) solid green 
curves is (0.35, 0.25) dex, respectively.

In Panel (D) of Fig.~\ref{fig:sims}, we examine the effect of hot environments. We impose an upper bound on the impact radius $d_{\rm impact}$ of 1, 2, 3$r_{\rm vir}$, where $d_{\rm impact}=r_{\rm vir}$ is equivalent to satellite quenching,
conventionally defined.
Simulations find that the impact of shock heating in massive halos is well beyond their virial radii, reaching roughly $3r_{\rm vir}$ \cite{2011bCen}. 
As expected,
a larger sphere of influence of hot halos gives rise to a larger reduction but the dependence is not strong.
To be clear, we also show the effect due to only the hot environment in crosses (labeled ``no self''), by removing only those halos that
are not already removed due to self-quenching
from the star-formation category.
Overall, the level of suppression by hot environments is comparable to, but somewhat less than, that due to self-heating.

The fact that the self-quenching and environment effects are closely intertwined is not surprising. It is due to mass segregation, especially at late times, where massive halos tend to reside in a hot environment. 
In other words, rich clusters of galaxies tend to 
contain a larger fraction of massive halos per unit 
mass of cluster than a less rich environment.
At $z=0$, approximately 1/3 self-heating halos are in a hot environment.

\begin{figure}
	\includegraphics[width=\columnwidth]{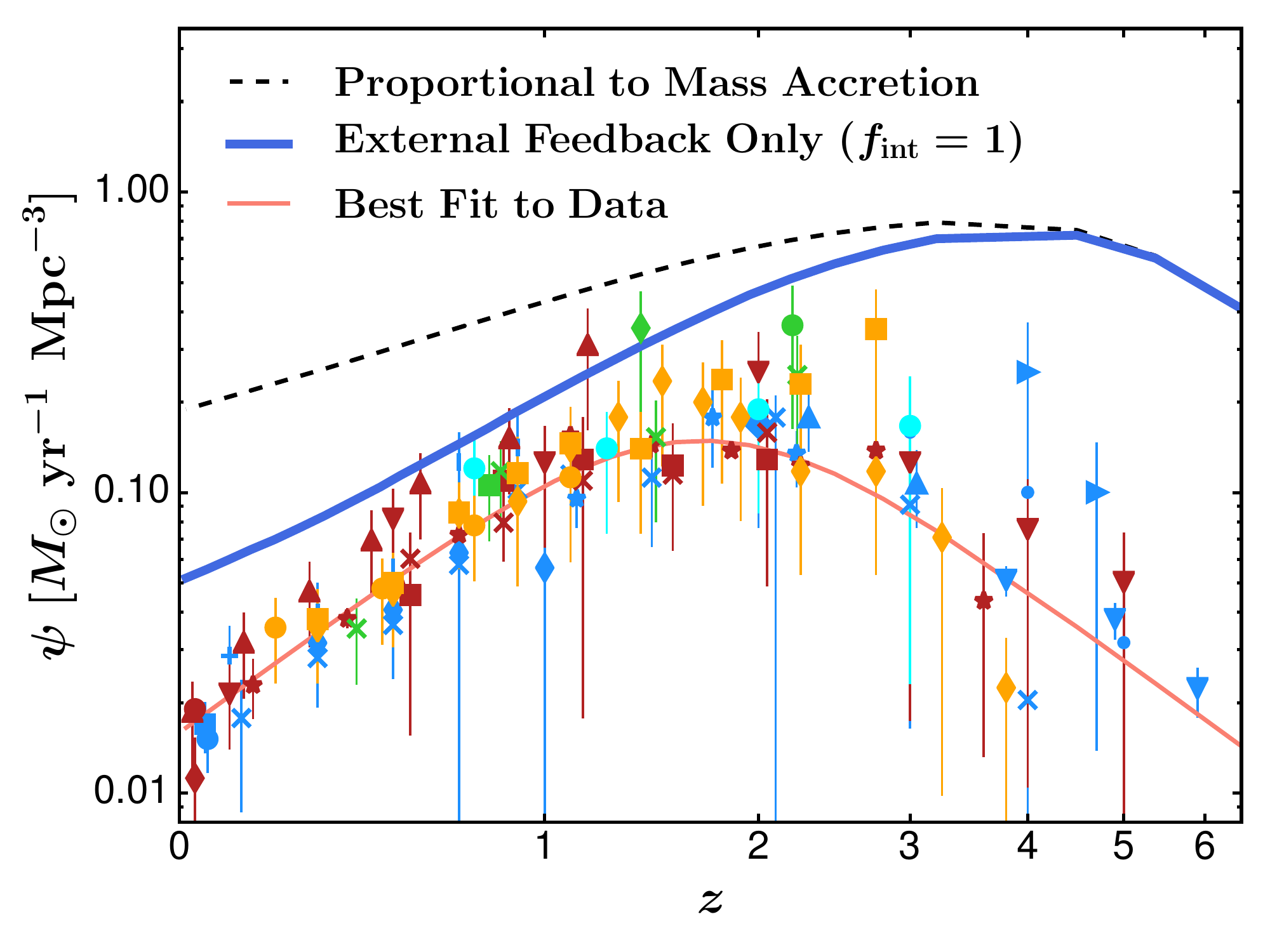}
	 \vspace{-0.2in}
	\caption{Comparison of our model (thick blue solid curve) to observational data in multiple bands---UV~(blue), IR~(red), H-$\alpha$~(green), UV+IR~(cyan), and 1.4 GHz~(orange). Refs. are listed in footnote~\ref{foot:data}. In the model, star formation is turned off in halos with $\sigma_v<20$ km/s (photoheating), $M_h > M_c$ (self-heating), and within $d=3r_{\rm vir}$ of a massive self-heating halo (hot environment). 	We also show a model without these cuts (black dashed curve) to illustrate the effect of mass accretion alone. The thin red solid curve is the best-fitting function to the data (Eq.~\ref{eq:fit}). Here we assume a Salpeter~\cite{1955Salpeter} initial mass function. 	}
    \label{fig:data}
\end{figure}

\subsection{Comparison to Data}
We next compare our model (with all three effects) to multi-wavelength observations in Fig.~\ref{fig:data}. We include compilations of ultraviolet~(UV), infrared~(IR), H-$\alpha$, and 1.4GHz data from Ref.~\cite{2013Behroozi} and Ref.~\cite{2014Madau}\footnote{\label{foot:data}
Data sources and their symbols in Fig.~\ref{fig:data}---
{\bf UV (blue)}:
$\CIRCLE$\cite{2005Wyder},
$\blacklozenge$\cite{2005Schiminovich},
$\blacksquare$\cite{2011Robotham},
$\times$\cite{2012Cucciati},
$\star$\cite{2007Dahlen},
$\blacktriangle$\cite{2009Reddy},
$\blacktriangledown$\cite{2012Bouwens,2012bBouwens},
$\blacktriangleleft$\cite{2013Schenker},
$\blacktriangleright$\cite{2006Yoshida},
$+$\cite{2007Salim},
$\Diamondblack$\cite{2011bLy},
$\bullet$\cite{2010vanderBurg},
$|$\cite{2007Zheng};
{\bf IR (red)}:
$\CIRCLE$\cite{2003Sanders},
$\blacklozenge$\cite{2003Takeuchi},
$\blacksquare$\cite{2011Magnelli},
$\times$\cite{2013Magnelli},
$\star$\cite{2013Gruppioni},
$\blacktriangle$\cite{2010Rujopakarn},
$\blacktriangledown$\cite{2009LeBorgne};
{\bf H-$\alpha$ (green)}:
$\CIRCLE$\cite{2011Tadaki},
$\blacklozenge$\cite{2009Shim},
$\blacksquare$\cite{2011Ly},
$\times$\cite{2013Sobral};
{\bf UV+IR (cyan)}:
$\CIRCLE$\cite{2010Kajisawa};
{\bf 1.4GHz (orange)}:
$\CIRCLE$\cite{2009Smolvic},
$\blacklozenge$\cite{2009Dunne},
$\blacksquare$\cite{2011Karim}.}.

In the model (blue solid curve), we cut off star formation in halos with $\sigma_v<20$ km/s (photoheating), $M_h>M_c$ (self-heating), or within 
$d_{\rm impact}=3r_{\rm vir}$ of a self-heating halo (hot environment). 
For comparison, we show the no-cut model assuming $\dot{M}_\star\propto \dot{M}_h$ and $f_{\rm int}=1$ (black dashed curve, same as in Fig.~\ref{fig:sims}). We also find the best-fitting function to data, 
\begin{align}\label{eq:fit}
\psi(z)=0.016\frac{(1+z)^{2.95}}{1+\left[(1+z)/2.70\right]^{5.91}} M_\odot \,{\rm yr}^{-1}\, Mpc^{-3},
\end{align}
shown as the thin red curve.

\begin{figure}
	\includegraphics[width=\columnwidth]{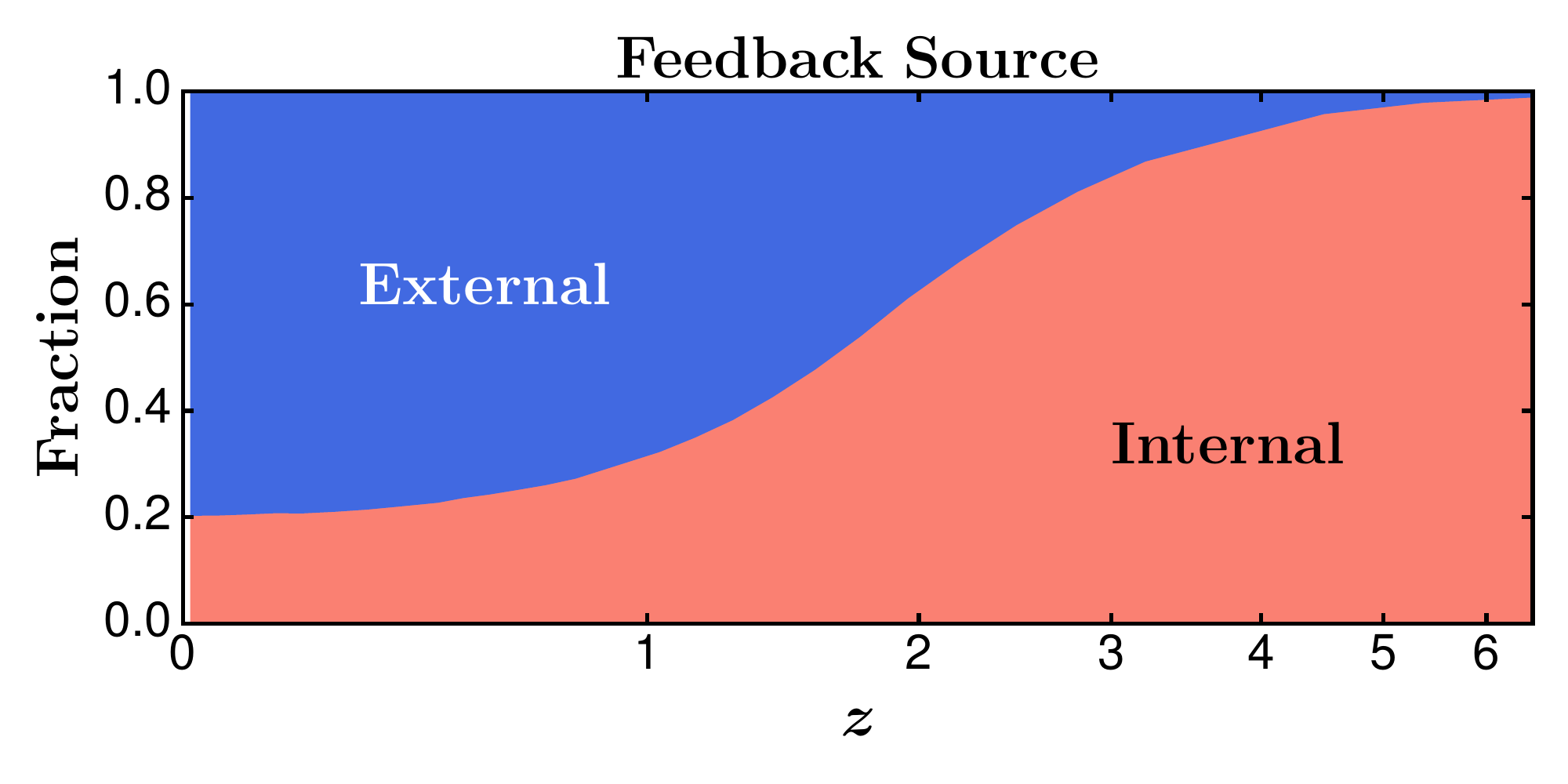}
	 \vspace{-0.2in}
    \caption{ The fractional contributions to the total negative feedback from external versus internal processes as a function of redshift. }
    \label{fig:ratio}
\end{figure}

Even though the mass accretion history shows a decreasing trend from $z=$~2--3 to $z=0$, it alone is too mild comparing to the data and therefore baryonic processes must 
have played an important role. 
At $z\le 2$, gravitational shock heating (including both self-heating and hot environments) 
can well explain the observed decline 
of SFR density.
To quantify the relative importance
of external heating induced versus
internally induced negative feedback processes as
a function of redshift, 
we show their respective fractional contributions as 
a function of redshift in Fig.~\ref{fig:ratio}.
The fraction due to the external heating process is defined to be 
ratio of the difference between the thin black dashed curve and the thick blue solid curve 
to the difference between 
the thin black dashed curve and the thin red solid curve;
the remainder is then designated as the fraction due to internal feedback processes.
It is seen that the strength of external feedback and that of internal feedback 
are roughly equal at $z\approx 1.7$.
Thus, at $z<1.7$ the external heating dominates
the suppression of star formation, whereas 
internal feedback does at $z>1.7$.
This is a new, fundamentally important result
because it indicates that the sharp decline of
SFR density from $z=2$ to $z=0$  can be primarily explained by the external feedback processes, and
does {\it not} requires major contribution from additional  internal feedback processes.

In contrast, at $z>2$, our model fails to match 
observations: the SFR density in our model continues to
rise to eventually peak at $z\approx 3$,
whereas the majority of observational data 
shows SFR density to peak at $z\approx 2$ followed by a continuous decline towards high redshift.
This tension may be alleviated if the current high redshift (mostly UV) observations
have significantly underestimated SFR density at early times. Alternatively, this is indicative of additional negative feedback from 
stellar evolution and/or AGN.

\begin{figure}
	\includegraphics[width=\columnwidth]{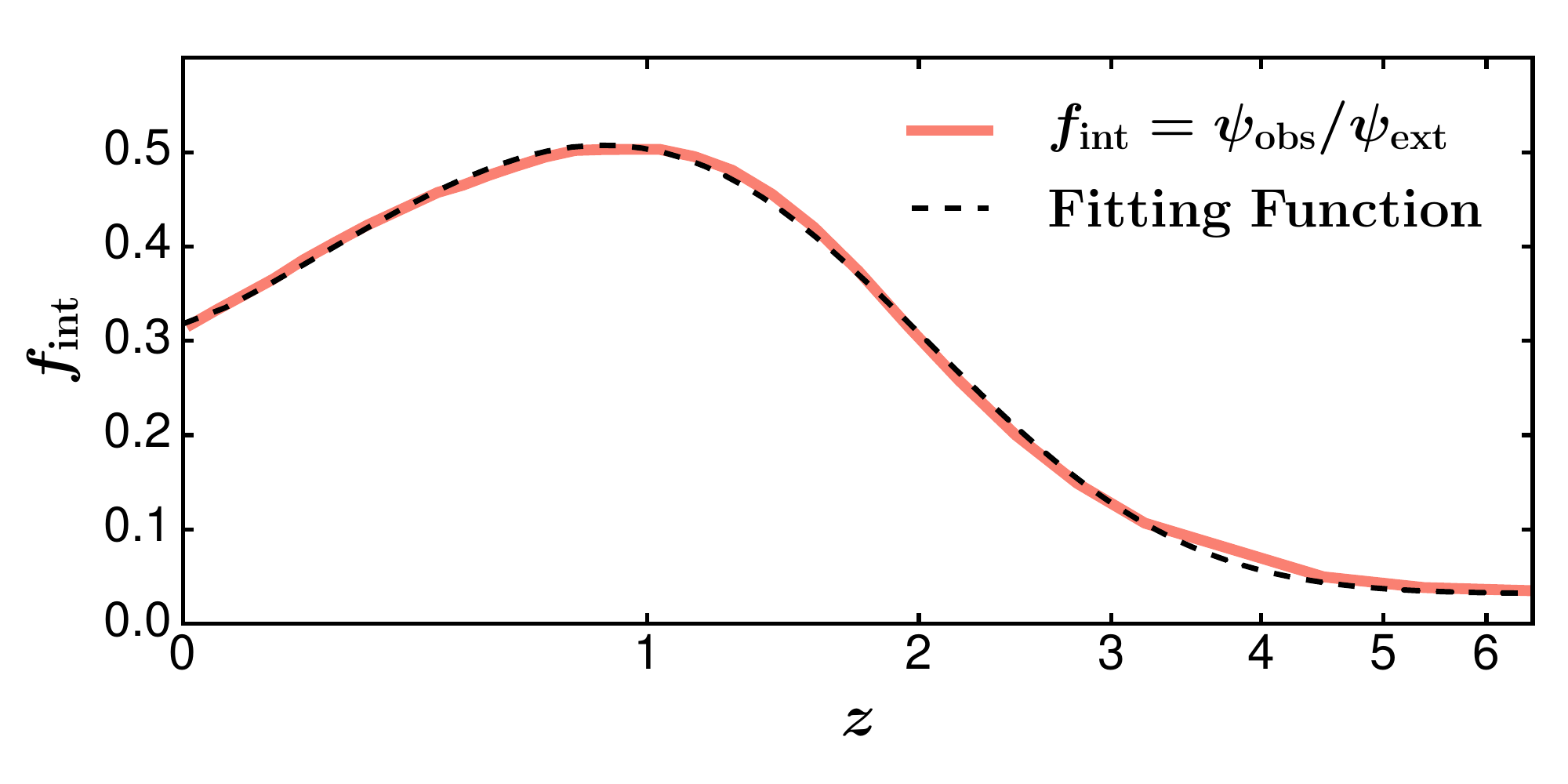}
	 \vspace{-0.2in}
    \caption{ {The internal star formation efficiency $f_{\rm int}$ needed to match our model (with only external feedback processes) to observations, i.e. the ratio between the thin red curve ($\psi_{\rm obs}$) and the thick blue curve ($\psi_{\rm ext}$) in Fig.~\ref{fig:data}.}}
    \label{fig:fint}
\end{figure}

Assuming that internal feedback processes are fully responsible for the difference between our model and data (i.e. the thick blue curve and the thin red curve in Fig.~\ref{fig:data}, respectively), we show in Fig.~\ref{fig:fint}  the required internal star formation efficiency $f_{\rm int}(z)$.
Our model can be implemented with N-body simulations
to address a range of issues concerning star formation
and its joint redshift and environment dependence, among others.
Our implementations are summarized in section~\ref{subsec:implementation}. We find the best fitting function for $f_{\rm int}$ (shown as the black dashed curve in Fig.~\ref{fig:fint}),
\begin{align}
f_{\rm int}(z) = 0.268\;(1+0.75z)^{6.81}\exp\left(-\frac{z^{0.922}}{0.302}\right)+0.032\,.
\end{align}

\begin{figure}
	\includegraphics[width=\columnwidth]{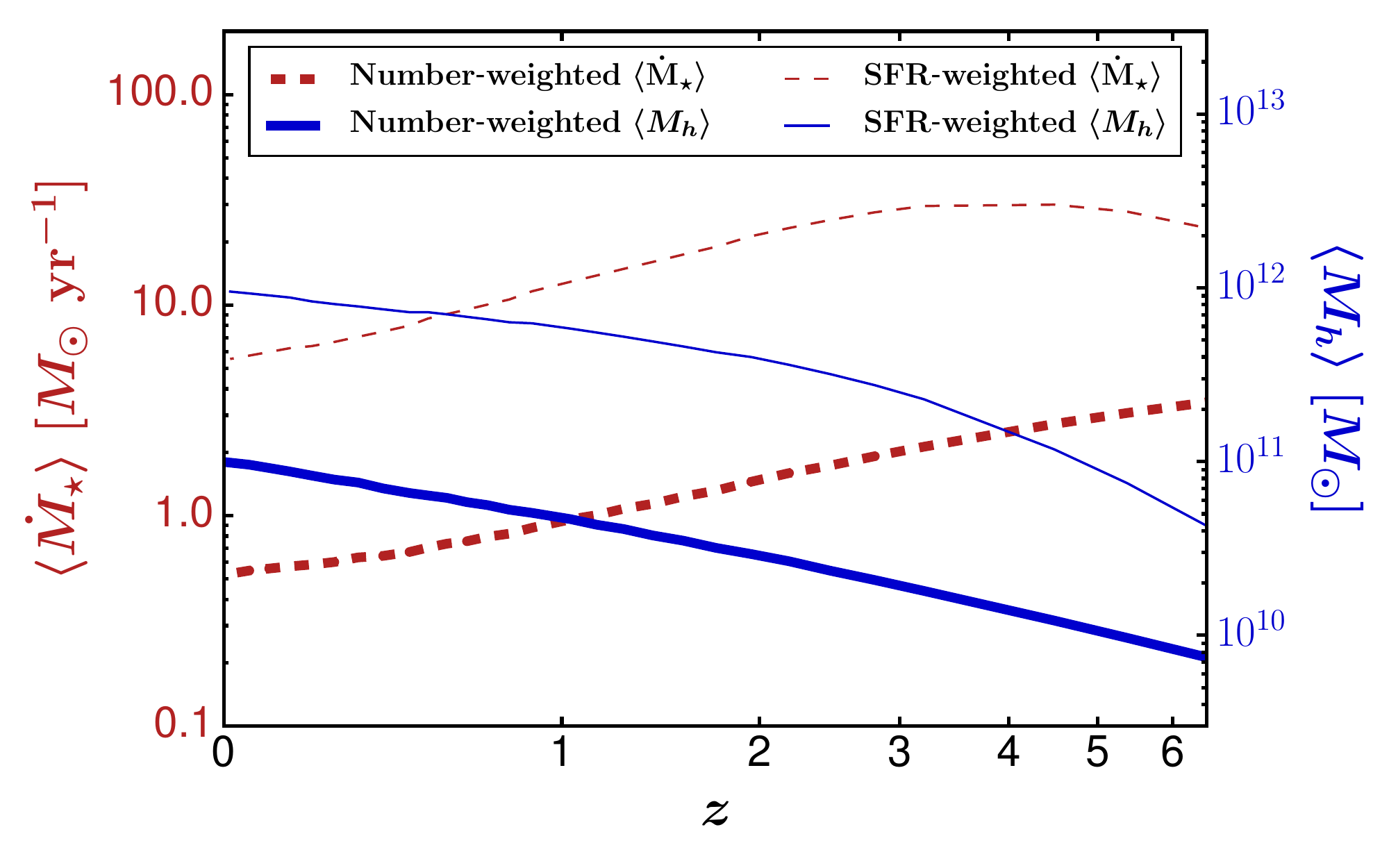}
	 \vspace{-0.2in}
    \caption{The average star formation rate $\langle \dot{M}_\star\rangle$ (red dashed curves, left axis) and the average halo mass $\langle M_h\rangle$ (blue solid curves, right axis) for star forming galaxies (with same cuts as in Fig.~\ref{fig:data}).  We show both the number-weighted (thick curves) and SFR-weighted (thin curves) average. Despite the ``downsizing'' trend in $\langle \dot{M}_\star\rangle$, the halos these galaxies reside in continue to grow in mass, showing no sign of declining. }
    \label{fig:down}
\end{figure}

Having addressed the issue of the evolution of cosmic SFR density, let us now turn to the issue
of so-called cosmic downsizing.
Observations find that 
the peak SFR of galaxies decreases
with decreasing redshift \cite{1996Cowie},
a trend that is in apparent opposition
to the hierarchical growth trend in the
standard cosmological model.
In Fig.~\ref{fig:down} we show the number-weighted and SFR-weighted average SFR $\langle \dot{M}_\star\rangle$ and the average halo mass $\langle M_h\rangle$ for star forming galaxies, with the same physical model corresponding to the solid blue curve in Fig.~\ref{fig:data}.
We show that with no freedom to adjust, our model predicts a downsizing trend in SFR
(the red dashed curves in
Fig.~\ref{fig:down})
that is in agreement with observations~\cite{1996Cowie,2013Magnelli}.
The average halo mass of these star forming galaxies is seen to continue to increase throughout the cosmic history, from $10^{11.5}M_\odot$ at $z$=2 to $10^{12}M_\odot$ at $z=0$ (SFR-weighted), showing no sign of ``downsizing''. 
It should be stressed that both the mean SFR
and the mean halo mass are based on star-forming halos, not all halos, at each redshift.
Nevertheless, this (star-forming) subset of halos shows an upsizing trend with decreasing redshift,
in tandem with the general hierarchical growth 
in the LCDM cosmological model.

To summarize,
we see no contradiction between the general
upsizing trend of halo mass expected in the
LCDM model
and the observed downsizing trend of SFR.
These two opposite trends
between SFR and halo mass with time
can be naturally understood:
the mass accretion rate at a given halo mass
decreases at a faster rate with decreasing redshift than the upsizing rate of the typical mass of star-forming halos.
In broad agreement with the conclusion reached
with respect to the SFR density evolution (Fig.~\ref{fig:data}),
the observed cosmic downsizing
at $z\le 2$ also does not require any additional significant negative feedback.

Taken together, our analysis suggests that 
the rapid downturn of SFR density and SFR in galaxies at $z\le 2$
may be mostly due to external feedback from gravitational shock heating,
whereas internal feedback from star formation and AGN may be required
to reconcile observations with theory at $z> 2$, if the present observational indications at 
$z>2$ hold up.

\section{Discussion}
\label{sec:discussion}

Proposals that advocate strong 
internal feedback processes to cause the sharp decline of star formation
and AGN activities at $z<2$ may face the conceptual difficulty that this recent redshift  is when the sources of the internal feedback, be it the star formation or the AGN activities, become the least vigorous.
Our finding that external feedback takes the leading role 
in suppressing star formation activities in this low redshift range
alleviates this logical difficulty.

On the other hand, the notion of strong, dominant internal feedback
required at $z>2$ is borne out in our analysis.
This outcome is aesthetically appealing for two reasons.
First, the bulk of galaxies at high redshift
are more moderate in gravitational potential well depths
and hence are significantly more prone to internal feedback 
processes. 
Second, as observed, star formation (say, gauged
by the specific star formation rate or alike)
and AGN activities themselves are more vigorous hence more negative feedback 
at high redshift.

One might run into another conceptual issue as to how
strong star formation activities and strong negative feedback may 
operate in a galaxy simultaneously.
We propose two ways.
First, in individual high redshift galaxies, the strong AGN/star formation episode and the consequent quiescent episode alternate, as the ejected intergalactic gas may cool to return to the galaxies, in the absence of external gravitational shock heating in massive halos.
The second way lies in a more global effort, where multiple generations of galaxies and galaxies in spatial proximity collectively and cumulatively contribute to raising the entropy of the circumgalactic and intergalactic medium~\cite{2015Cen}. Over time, the net gas accretion onto galaxies is retarded and reduced. 
In some cases, such heating is long-lasting. As a particular example, heating of the circumgalactic and intergalactic medium
by internal processes at their prime times (i.e., $z>2$) 
may continue to affect gas accretion onto galaxies at later times $z<2$ 
when internal feedback has much diminished, in conjunction with external heating by gravitational shocks.

Overall the entire redshift range, the question of 
the relative importance between supernova feedback 
and AGN feedback is open.
What has become increasingly clear is that most large-scale cosmological
hydrodynamic simulations have significantly under-estimated
the effects of supernova feedback due to inadequate numerical implementations.
Ref.~\cite{2014Kimm} found that momentum injection due to supernova feedback may be significantly underestimated at the typical resolutions employed by current large-scale cosmological hydrodynamic simulations~\cite{2011bCen}.
In simulations with supernova feedback properly implemented, 
there is strong evidence that the overcooling and
stellar overproduction problem is rectified in 
halos as massive as $10^{11.5}M_\odot$ at $z=3$ \cite{2015Kimm}.
Thus, it remains unclear if a large amount of AGN feedback,
which may be a symptom of an under-estimation of supernova feedback,
is still necessary or perhaps should be avoided.
This is not to reject the notion that AGN feedback may be important for some
subset of systems, such as 
the central galaxies in clusters of galaxies, where
AGN feedback, in combination with other 
processes (conduction, gravity waves, etc.),
may play a significant role to 
periodically disrupt, suppress, or retard
cooling flows.

For central galaxies hosted by massive halos, the dense gas core 
may cool rapidly via X-ray emission, forming the so-called ``cooling flow''~\cite{Fabian1994}. Star formation has indeed been observed in some of these galaxies~\cite{McDonald2011}. To test the impact of the cooling flow, we let the gas in the central regions ($< 0.1r_{\rm vir}$, following Ref.~\cite{Gaspari2017} of all massive ($M>M_c$) halos to form stars. We find that the gas accreted onto this central region amounts to only $\approx1\%$ of the total accreted gas, and hence it has negligible contribution to the overall star formation rate.

\section{Conclusions}
\label{sec:conclusions}
We perform a joint analysis of the evolution 
of the global star formation rate density \cite{1996Madau,1998Madau} 
and the observed ``downsizing'' phenomenon~\cite{1996Cowie}
in the redshift range $z=0$--6,
in the context of the standard LCDM model.
We implement the external,
 star-formation suppression effects of two important known physical processes--- photoheating due to reionization of the intergalactic medium and gravitational shock heating due to formation of massive halos and large-scale structure--- utilizing the accurate halo catalogues from the Bolshoi simulation.

We show that, at $z\le 2$, gravitational shock heating, including self-heating of massive halos and hot environments, can well explain both the observed decline in SFR density 
(Fig.~\ref{fig:data}) and
the ``downsizing'' trend in SFR
(Fig.~\ref{fig:down}).
We find a comparable level of impact from self-heating and hot environments. These two effects are significantly 
entangled due to halo mass segregation.
We also find that the typical halo mass of star forming galaxies,
which are a {\it subset} of all halos,
steadily increases from $z=2$ to $z=0$, in tandem with the hierarchical structure formation picture in the LCDM model.

The photoheating effect is found to play only a minor role in suppressing star formation, mostly at high redshift.
The combined effect of gravitational shock heating and photoheating appears insufficient, and additional negative feedback effects
are required to reconcile with observations at $z>2$.
Internal feedback effects from stellar evolution and supermassive black hole growth are natural candidates for this role.
This apparent requirement at $z>2$ is physically attainable 
and logically more self-consistent,
because galaxies at $z>2$ are more moderate in mass and stronger in star formation (i.e., much higher specific star formation rates
or specific AGN rates),  
thus allowing for stronger negative feedback.
In terms of suppressing star formation globally, 
the gravitational shock heating dominates at $z<1.7$,
where internal feedback 
dominates at $z>1.7$.
Fig.~\ref{fig:ratio} summarizes this finding.

The overall picture laid out here would
relieve the need of a seemingly bewildering notion that
star formation or AGN activities quench themselves permanently. 
Without strong external quenching, a quiescent episode resulting from internal feedback would inevitably be 
followed by a new episode of gas cooling and star formation. 
Quenching by strong internal feedback can only be temporary. 
Rather, our work supports the notion that cold gas supply
determines the overall star formation and AGN activities.
Shutoff of cold gas supply is a necessary condition
for a persistent, long-term quenching of star formation.
External feedback, i.e., gravitational shock heating of the gas to be accreted,
provides the desired solution.

\section{Acknowledgement}
We thank the Bolshoi collaboration \cite{2016Klypin}
and Rockstar collaboration \cite{2013Behroozi, 2013bBehroozi} for providing 
the Bolshoi-Planck simulation catalogs.
We thank Greg Bryan, Zoltan Haiman, Jerry Ostriker, David Spergel, David Weinberg for useful discussions. The analysis is in part performed at the TIGRESS high performance computer center at Princeton University. 
JL is supported by an NSF Astronomy and Astrophysics Postdoctoral Fellowship under award AST1602663. 
This work is supported in part by grants NNX12AF91G and AST1515389. 

\bibliographystyle{physrev}
\bibliography{paper}
\end{document}